\documentclass[acus]{JAC2000}

\usepackage{graphicx}

\newcommand{\be}{\begin{equation}}
\newcommand{\ee}{\end{equation}}
\newcommand{\bea}{\begin{eqnarray}}
\newcommand{\eea}{\end{eqnarray}}

\begin{document}
\title{\centering QCD sum rule analysis of 
{\boldmath $V$} and {\boldmath $A$} current correlators
from {\boldmath $\tau$}-decay data}

\author{Konstantin N. Zyablyuk, ITEP,  Moscow, Russia}

\maketitle

\begin{abstract}
2-point correlators of vector and axial currents, obtained from 
$\tau$-decay data, are studied
within the framework of perturbative QCD and Operator Product Expansion.
Various sum rules, obtained from 
Borel transformation of the correlators in complex plane,
are used to separate the contributions of different operators 
from each other. The analysis confirms the $Q^2$-dependence
of the correlators in the space-like region, predicted
by QCD+OPE. However the operator values are found to be in 
certain disagreement with the estimations, obtained from other data.
\end{abstract}

\section{Objectives}

Precise measurements of vector $V$ and axial $A$ spectral functions
in hadronic $\tau$-decays by ALEPH \cite{ALEPH2} and OPAL \cite{OPAL}
collaborations provide us with possibility to test
various QCD aspects. Perturbation Theory (PT) and 
Operator Product Expansion (OPE) are the most well-established ones.
Here we shall compare theoretical predictions with the data 
within the framework of sum rules. Particular details of this analysis 
can be found in \cite{IZ, GIZ}. 

The 2-point correlators of charged vector and axial-vector 
currents
$$
J=V,A: \qquad V_\mu=\bar{u}\gamma_\mu d \, , \quad 
A_\mu=\bar{u}\gamma_\mu\gamma_5 d 
$$
can be parametrized by 2 polarization functions $\Pi(q^2)$: 
$$
i\int e^{iqx}\left< J_\mu(x) J_\nu(0)^\dagger \right> dx = \hspace{20mm} 
$$
\be
=(q_\mu q_\nu -g_{\mu\nu}q^2)\Pi_J^{(1)}(q^2)+q_\mu q_\nu\Pi_J^{(0)}(q^2)
\ee
For $q^2=s>0$ they have imaginary parts, the so-called spectral functions
\be
v_1/a_1(s)=2\pi\, {\rm Im}\,\Pi_{V/A}^{(1)}(s+i0)
\ee
which have been measured from hadronic $\tau$-decays for $0<s<m_\tau^2$,
the plots can be found in \cite{ALEPH2, OPAL}. The scalar axial 
polarization function $\Pi_A^{(0)}$
is basically saturated by single pion decay channel.
Its imaginary part $a_0$ is delta-function, which can be easily 
separated from  $a_1/v_1$.

It turns out to be convenient to consider the sum and 
difference $v_1\pm a_1$ instead of $v_1$ and $a_1$ separately. Indeed,
the sum $v_1+a_1$ is known with better accuracy, while the difference 
$v_1-a_1$ does not contain perturbative terms in
the massless quark limit. The QCD expressions for appropriate polarization
functions can be written in the following form:
\bea
\Pi_V^{(1)}(s)-\Pi_A^{(1)}(s) & = & \sum_{k\ge 2}{O_{2k}^{V-A}\over (-s)^k}
\label{pivma}\\
\Pi_V^{(1)}(s)+\Pi_A^{(1)}(s) & = & -{1\over 2\pi^2}\ln{-s\over \mu^2}+
\mbox{higher loops} \nonumber \\
 & & + \sum_{k\ge 2}{O_{2k}^{V+A}\over (-s)^k}  \label{pivpa}
\eea
The $2k$-dimensional constants $O_{2k}^{V,A}$ are the vacuum expectation
values of the operators, constructed from the quark and 
gluon fields \cite{SVZ}.
They have been computed up to dimension $D=8$. The numerical values 
of $O_{2k}$ cannot be determined within the perturbation theory. 

\begin{figure}[b]
\centering
\includegraphics[width=70mm]{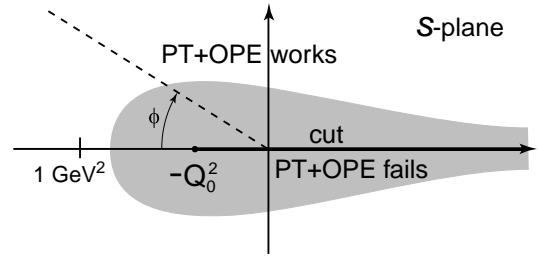}
\caption{Region of validity of perturbation theory
and operator product expansion} \label{pt_ope}
\end{figure}

Obviously the expressions (\ref{pivma},\ref{pivpa}) are not valid for
all values of $s$. Exact polarization operator $\Pi(q^2)$ is known
to be an analytical function of $s=q^2$ with a cut along positive real 
semiaxes. So it is convenient to study the QCD predictions   
(\ref{pivma},\ref{pivpa}) in the whole complex $s$-plane. 
These series are not valid for small $|s|$, where effective degrees
of freedom are hadrons rather than quarks. Moreover, the higher
loop perturbative terms in (\ref{pivpa}) have unphysical cut starting
from some $s=-Q_0^2<0$.  The OPE series
with finite number of operators does not have a cut along positive real
semiaxis, but has very singular behavior at $s=0$. Based upon these
speculations one may draw schematic Figure \ref{pt_ope}, 
displaying the region
of validity of the series (\ref{pivma},\ref{pivpa}).

Another drawback of QCD is that these series
are likely to be asymptotic, i.e. divergent for any fixed $s$. 
The way to deal with divergent series is to apply Borel transformation
\be
{\cal B}_{M^2}\Pi = \mbox{pert. terms}+
\sum_{k\ge 2}{O_{2k}\over (k-1)!\, M^{2k}}
\label{bor0}
\ee
which improves the convergence by suppressing the higher terms. 
It is not clear, whether it improves the perturbative
series, which is an expansion in inverse powers of $\ln{(-s)}$, rather
than $s$ itself. However the expansion in $1/\ln{M^2}$ might be convergent.

The primary goal of the investigation is to find the numerical values
of input theoretical parameters, such as $\alpha_s(m_\tau^2)$ and few 
operators of lowest dimensions. We shall compare the QCD result for
the Borel transform (\ref{bor0}) of the series (\ref{pivma},\ref{pivpa})
with the experimental values, computed by exploiting the analytical
properties of exact polarization functions. In order to separate the
operators from each other, we shall consider the Borel transform 
(\ref{bor0}) at complex values of the argument $M^2 e^{i\phi}$.
This can be alternatively understood as the Borel transformation
applied to the polarization function, taken at the angle $\phi$
w.r.t. the real negative semiaxes in the $s$-plane, see Fig.~\ref{pt_ope}.
We shall also try to find the lowest value of the Borel mass $M^2$, 
at which the comparison of QCD to the experiment can be made. 

\section{{\boldmath V-A} sum rules}

We start the analysis from the $V-A$ case (\ref{pivma}) which is purely
nonperturbative. The dispersion relation for the difference of 
polarization functions does not need subtractions and is written
in the following way:
$$
\Pi_V^{(1)}(s)-\Pi_A^{(1)}(s)= \hspace{20mm}
$$
\be
={1\over 2\pi^2}\int_0^\infty {(v_1-a_1)(s')\over s'-s}ds' + {f_\pi^2\over s}
\label{vmadr}
\ee
The last term is the kinematic pole which is specific feature of
axial currents. Indeed, the r.h.s. has appropriate asymptotics:
at $s\to 0$ it matches the chiral theory prediction, while the 
expansion at $s\to \infty$ starts from the operator of dimension $D=4$,
as it should be. 

Applying the Borel transformation to (\ref{vmadr}), one gets the following 
sum rule:
$$
\int_0^\infty e^{-s/M^2} (v_1-a_1)(s) {ds\over 2\pi^2} =
$$
\be
=f_\pi^2 + \sum_{k\ge 2} {O_{2k}^{V-A}\over (k-1)! \, M^{2k-2}}
\label{bor0vma}
\ee
One may estimate the numerical values of the operators $O^{V-A}$ up to 
dimension 8 from other data:
\bea
O_4^{V-A} & = & 2(m_u+m_d)\left<\bar{q}q\right> = - f_\pi^2 m_\pi^2
\nonumber \\
 & & \mbox{negligible at} \; s\sim 1 \, {\rm GeV}^2,
\nonumber \\
O_6^{V-A} & = & - {64\over 9}\pi\alpha_s\left<\bar{q}q\right>^2
\approx -2\times 10^{-3}\,{\rm GeV}^6,
\nonumber \\
O_8^{V-A} & = & 8\pi\alpha_s m_0^2\left<\bar{q}q\right>^2
\approx 2\times 10^{-3}\, {\rm GeV}^8,
\label{opvma} 
\eea
where
\be
m_0^2 = {\left<q\hat{G}q\right>\over i\left<\bar{q}q\right>} =
0.8\pm 0.2 \, {\rm GeV}^2
\label{m02}
\ee
has been found from barionic sum rules \cite{BI}. In the numerical 
estimation we assumed $m_u+m_d=12 \, {\rm MeV}$ and $\alpha_s=0.5$ 
at 1 GeV${}^2$. The factorization hypothesis was used in order to 
bring the operators $O_{6,8}$ to the form (\ref{opvma}). It has internal
theoretical ambiguity $\sim 1/N_c^2$ among the $D=8$ operators, 
see \cite{IZ}.  

QCD corrections to the operators
$O_6$ have been computed in \cite{AC}. They turn out to be large and
may increase the effective contribution of the $D=6$ operator by
about 50\%:
\bea
O_6^{V-A}& =&  - {64\over 9}\pi\alpha_s\left<\bar{q}q\right>^2
\left[1+{\alpha_s\over \pi}\left({1\over 4}\ln{-s\over\mu^2}+c_6\right)
\right] \nonumber \\
 & \approx &  -3\times 10^{-3}\,{\rm GeV}^6
\label{o6num}
\eea
The coefficient $c_6$ is ambiguous:
two essentially different choices  were presented in \cite{AC}. In the 
numerical estimation (\ref{o6num}) we used more moderate one $c_6=89/48$.

One sees, that the r.h.s. of (\ref{bor0vma}) has leading term $f_\pi^2$
and relatively small (but interesting) contributions of $O_{6,8}$ at
$M^2>0.5\,{\rm GeV}^2$. One way to kill $f_\pi^2$ is to differentiate 
(\ref{bor0vma}) by $M^2$. This however inevitably increases the errors
of the experimental integral. It seems more effective to perform
another trick: one substitutes complex Borel mass $M^2e^{i\phi}$ into
(\ref{bor0vma}) and takes imaginary part of it. The result is:
$$
\int_0^\infty  e^{-{s\over M^2}\cos{\phi}}
\sin{\left({s\over M^2}\sin{\phi}
\right)}(v_1-a_1)(s){ds\over 2\pi^2 M^2}
$$
\be
=\,-\sum_{k\ge 2}{\sin{\left((k-1)\phi\right)}\over (k-1)!}\,
{O_{2k}^{V-A}\over M^{2k}}
\label{bor1vma}
\ee     

\begin{figure*}[tb]
\centering
\includegraphics[width=60mm]{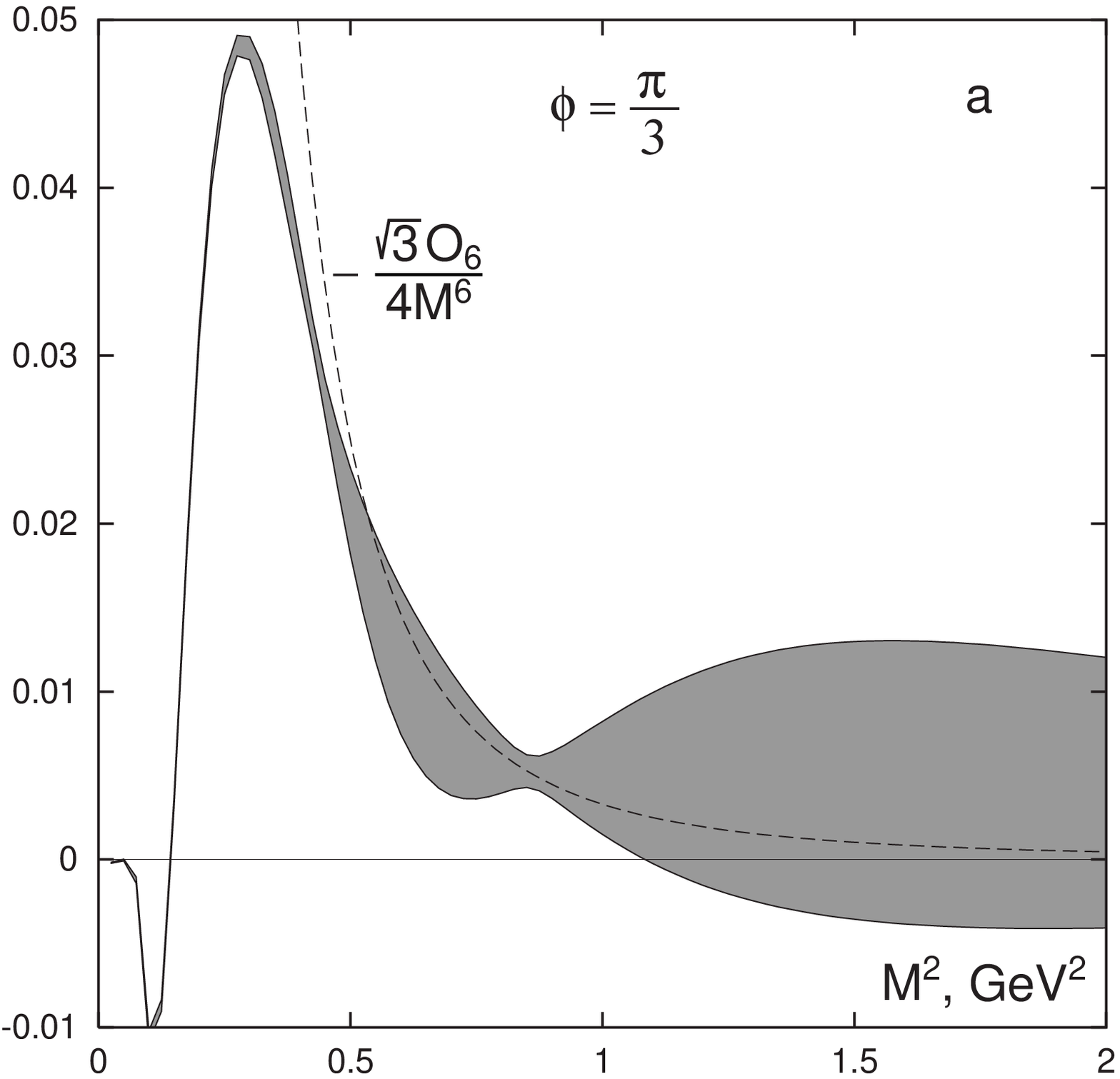} \hspace{25mm}
\includegraphics[width=61mm]{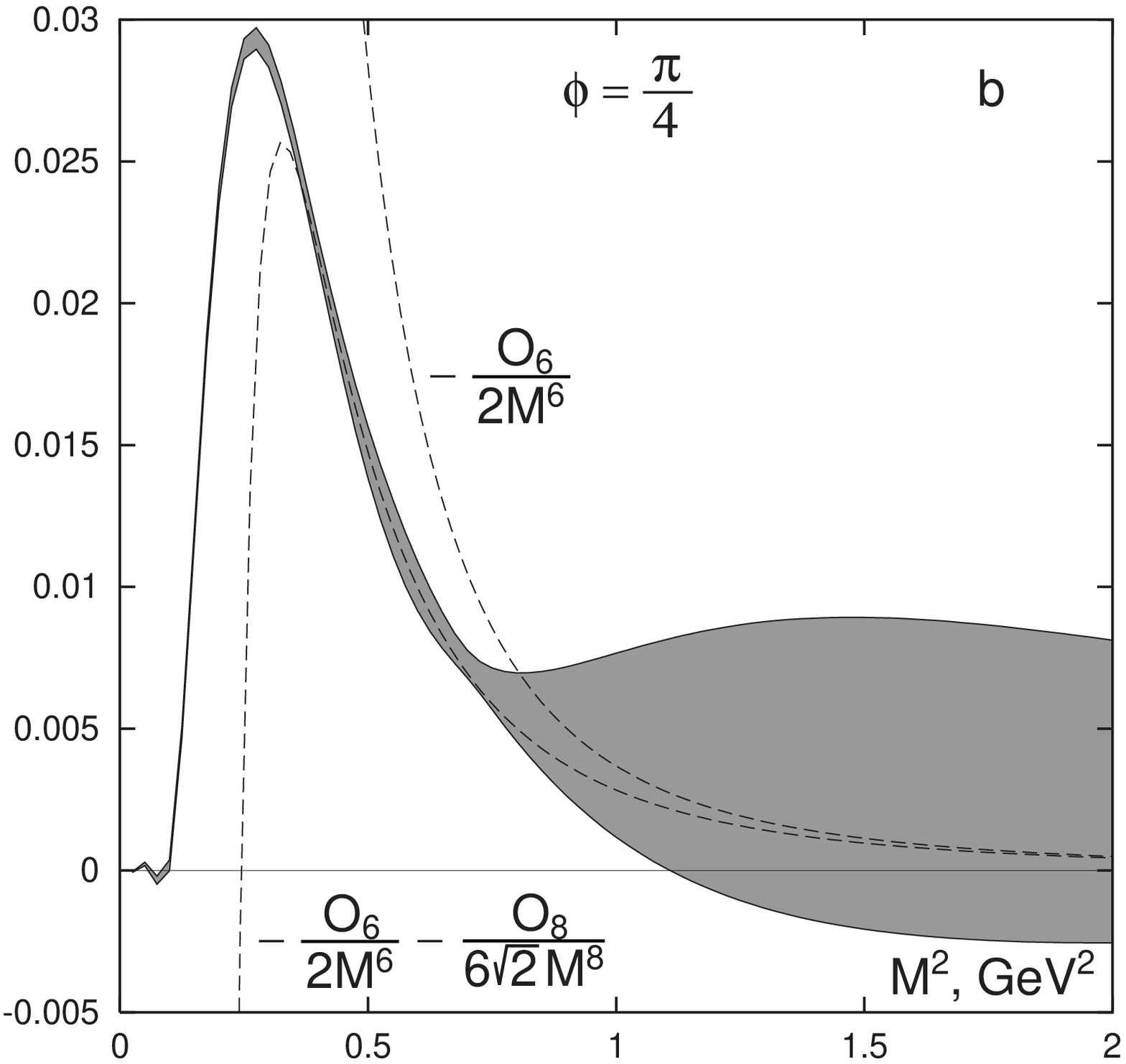}
\caption{Sum rule (\ref{bor1vma})
for $\phi=\pi/3$  (a) and $\phi=\pi/4$ (b). Dash lines display OPE
prediction with operators (\ref{vmafit}).} \label{sr1vma}
\end{figure*} 

Let us consider the angle $\phi=\pi/3$. The operator $O_8$ disappears from 
the r.h.s. of (\ref{bor1vma}) and only $O_6$ is important in this case. 
The l.h.s. of (\ref{bor1vma}) 
is shown in Fig.~\ref{sr1vma}a as shaded area (the upper integration 
limit is $m_\tau^2$, since there are no data beyond this point). The
errors have local minimum at the point $M^2=0.8\,{\rm GeV}^2$. It happens
because the $\sin{(\ldots)}$ in the integral has zero at $s=m_\tau^2$ and
thereby suppresses large experimental errors. At this point we determine
the operator $O_6^{V-A}$ and plot the r.h.s. of (\ref{bor1vma}) with
this value in Fig.~\ref{sr1vma}a as dash line. 

Second interesting angle is $\phi=\pi/4$. Both $O_6$ and $O_8$ contribute,
but the next term with $O_{10}$ disappears. This means that one may 
go to lower values of $M^2$ in order to reduce the experimental uncertainty.
Indeed, as can be seen from Fig.~\ref{sr1vma}b, the agreement can be
achieved down to $M^2=0.4\,{\rm GeV}^2$ in this case. At this point
we obtain the most accurate value of the operator $O_8$. 

The result of the fit:
\bea
O_6^{V-A} & = & -(6.8\pm 2.1)\times 10^{-3} \,{\rm GeV}^6 \nonumber \\ 
O_8^{V-A} & = & (7\pm 4)\times 10^{-3} \,{\rm GeV}^8   
\label{vmafit}
\eea
(details of the fit and error estimations are discussed in \cite{IZ}).
The result for $O_6$ is twice larger than our estimation (\ref{o6num}).
It might have different explanations: overestimated $m_u+m_d$,
failure of factorization, large $\alpha_s$ corrections are the few ones
among them. But the mass $m_0^2$, obtained from (\ref{vmafit}) is in agreement
with (\ref{m02}).

\section{Perturbative series}

\begin{figure}[b]
\centering
\includegraphics[width=82mm]{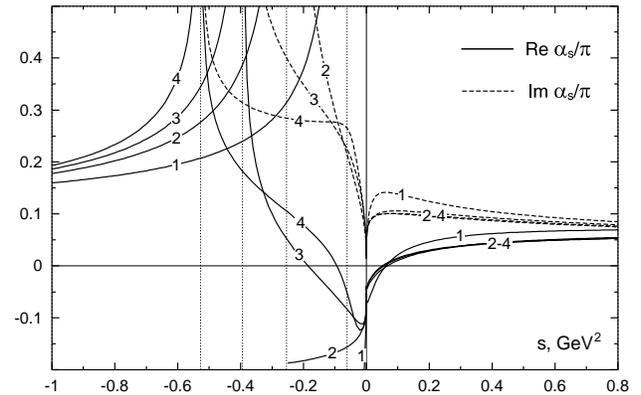}
\caption{Real and imaginary parts of 
$\alpha_{\overline{\rm MS}}(s)/\pi$  as
exact numerical solution of RG equation (\ref{rge}) on real axes for 
different number of loops. The initial condition is chosen
$\alpha_s=0.355$ at $s=-m_\tau^2$. Vertical dotted lines display the
position of the unphysical singularity at $s=-Q_0^2$ for 
each approximation ($4\to 1$ from left to right).}
\label{alpha}
\end{figure}

Before analyzing $V+A$ sum rules, we outline the basic features of
perturbative series. The QCD coupling $a\equiv \alpha_s/\pi$
is a function of the scale $Q^2$, determined by the renormalization
group equation:
\be
{d a\over d\ln{Q^2}}=-\beta(a)=-\sum_{k\ge 0} \beta_k a^{k+2}
\label{rge}
\ee
where the factors $\beta_k$ have been computed up to 4 loops
in $\overline{\rm MS}$ scheme \cite{RVL}. In particular 
$\beta_0=4/9$, $\beta_1=4$, $\beta_2=10.06$ and $\beta_3=47.23$
for 3 flavors. The solution of RG equation is
\be
\ln{Q^2\over \mu^2}=-\int_{a(\mu^2)}^{a(Q^2)} {da\over\beta{(a)}} \; ,
\qquad Q^2=-s
\ee
Since the integral is convergent at $\infty$ for any fixed order
(at least with positive $\beta_k$),
the coupling $a(s)$ has unphysical singularity at some negative 
$s=-Q_0^2$, see Fig.~\ref{pt_ope}. The 
properties of the solution of RG equation 
can be understood by viewing on Fig.~\ref{alpha}.
 
The polarization function is obtained by integrating the Adler
function, which is finite and has been computed up to N${}^3$LO term
in $\overline{\rm MS}$ \cite{SS}:
$$
D(Q^2)=-2\pi^2{d\Pi (Q^2)\over d\ln{Q^2}} \hspace{20mm}
$$
\be
=1+a+K_2a^2+K_3a^3+{\rm unknown}
\label{adler}
\ee
where $K_2=1.64$ and $K_3=6.37$ for 3 flavors. In our calculations 
we shall take the theoretical uncertainty equal to the contribution 
of the last term in (\ref{adler}), $\pm K_3 a^3$. Since we do not know
$K_4$, it is reasonable to use only 3-loop approximation also for the
$\beta$-function in (\ref{rge}).

The polarization function constructed in this way has unphysical
cut from $s=-Q_0^2$ to $s=0$. It is an obvious indication of QCD 
inapplicability at low $|s|$. However there are certain attempts
to construct the perturbative functions with appropriate analytical
properties on the whole $s$-plane, for instance by constructing
an analytical QCD coupling with help of dispersion relation \cite{ShSol1}
(subtractions assumed):
$$
\alpha_s(s)_{\rm an}={1\over \pi}\int_0^\infty { {\rm Im}\,
\alpha_s(s')\over s'-s}ds' \hspace{5mm}
$$
\be
={\pi\over\beta_0}\left( {1\over \ln{(-s/\Lambda^2)}}-{\Lambda^2\over
\Lambda^2+s}\right)+\ldots
\label{alpan}
\ee
This way is not unique: one may write down the same dispersion relation
for the polarization function $\Pi(s)_{\rm an}$ as well. At the NLO level the
result will be the same as the substitution of (\ref{alpan})
into (\ref{adler}), but it is not the case for higher terms. In general,
the purely logarithmic terms in analytic approach are the same as
in conventional QCD, but the power terms are different, and there
appears $D=2$-like term $\sim 1/s$, absent in canonical OPE. 

\begin{figure}[t]
\centering
\includegraphics[width=82mm]{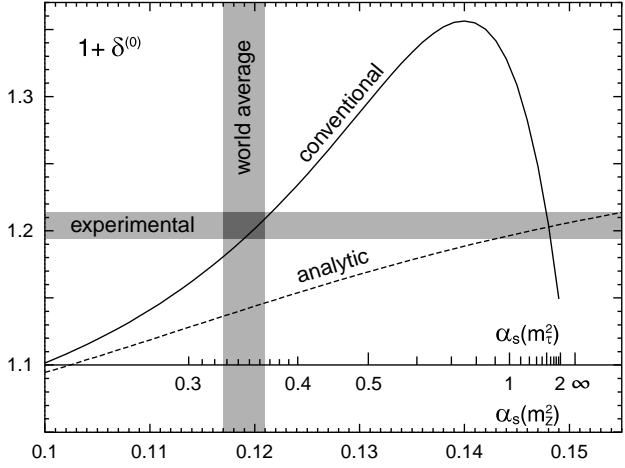}
\caption{Correction $\delta^{(0)}$ versus $\alpha_s(m_\tau^2)$ and
$\alpha_s(m_Z^2)$ in conventional and analytic approach in 
3-loop approximation.} 
\label{d0_fig}
\end{figure}

QCD must give correct value of the hadronic $\tau$-decay branching 
ration $R_\tau\sim 3(1+\delta^{(0)})$, which is measured with
rather high accuracy. It is also weakly sensitive to the 
nonperturbative power corrections. The perturbative fractional correction
$\delta^{(0)}$ is given by well-known formula (e.g. \cite{BNP}):
$$
1+\delta^{(0)}=1.206\pm 0.010 \hspace{35mm} 
$$
\be
=2\pi i \oint_{|s|=m_\tau^2}
 {ds\over m_\tau^2}\left(1-{s\over m_\tau^2}\right)^2
\left(1+2{s\over m_\tau^2}\right)\Pi(s)
\label{delta0}
\ee
$$
\mbox{where} \qquad \Pi=\Pi_V^{(1)}+\Pi_A^{(1+0)} \,  .
$$
Notice, that the circle integral includes the contribution of
unphysical cut, while in any analytical approach it is thrown away.
The numerical results for (\ref{delta0}) in both approaches are shown
in Fig.~\ref{d0_fig}. It is seen, that analytical scheme predicts
very large $\alpha_s(m_Z^2)$ (at $Q^2=m_Z^2$ the difference between
both approaches is not important) and, therefore, fails to agree with
other data. So we shall not consider it anymore.
 
It follows from Fig.~\ref{d0_fig}:
\be
\alpha_s(m_\tau^2)=0.355\pm 0.025
\label{alpha_f}
\ee
The error includes the theoretical uncertainty $\pm K_3a^3$ of Adler
function. This result is 3-loop, the 4-loop result with the estimation
$K_4=25-50$ would give us slightly less value within the error range. 
We notice also, that there is second point on Fig.~\ref{d0_fig} where 
the conventional curve crosses the experimental band. However it is
unstable under changes of various perturbative input parameters and
prescriptions and cannot be considered as reliable one.

\section{{\boldmath $V+A$} sum rules}

The operators $O^{V+A}$ in (\ref{pivpa})
 include purely gluonic condensates. In particular,
\be
O_4^{V+A}={\alpha_s\over 6\pi}\left<G_{\mu\nu}^a G_{\mu\nu}^a\right>
\ee
The $D=4$ gluonic condensate has been found from charmonium sum rules
\cite{SVZ}:
$$
\left< {\alpha_s\over \pi} G^2\right> = 0.012 \, {\rm GeV}^4
\qquad  \eqno{({\rm SVZ})}
$$
The $D=6$ operator contains gluonic condensate $\sim \left< G^3 \right>$,
which is not known. The quark contribution after factorization get the form:
\be
O_6^{V+A}={128\over 81}\pi\alpha_s \left< \bar{q}q\right>^2
= (1.3\pm 0.5)\times 10^{-3}\,{\rm GeV}^6
\label{o6vpa}
\ee
For numerical estimation we used our $V-A$ fit (\ref{vmafit}) and added
additional error which might occur due to incomplete 
cancellation of two relatively large term in the sum $V+A$ after
factorization.  The $D=8$ operator cannot be obtained from other data,
but we estimate its upper limit as $|O_8^{V+A}|<10^{-3}\,{\rm GeV}^8$.
Details given in \cite{GIZ}. So $O_{6,8}^{V+A}$ are essentially 
smaller than $O_{6,8}^{V-A}$ and perturbative terms dominate here.

Now let us define the Borel transform of the polarization function
$\Pi_{V+A}$:
$$
B_{\rm exp}(M^2)=\int_0^{m_\tau^2}e^{-s/M^2}(v_1+a_1+a_0)(s){ds\over M^2}
$$
\be
= B_{\rm pt}(M^2)+2\pi^2 \sum_{k\ge 2}{O_{2k}^{V+A}\over (k-1)!\, M^{2k}}
\label{borvpa}
\ee
The perturbative part $B_{\rm pt}$ is computed numerically:
$$
B_{\rm pt}(M^2)=i\pi \oint e^{-s/M^2} \Pi_{\rm pt}(s){ds\over M^2}
$$
The integration contour goes counterclockwise from $s=m_\tau^2+i0$ to
$s=m_\tau^2-i0$ around the cut of the perturbative polarization function
$\Pi_{\rm pt}$ (including unphysical part). 

\begin{figure*}[t]
\centering
\includegraphics[width=80mm]{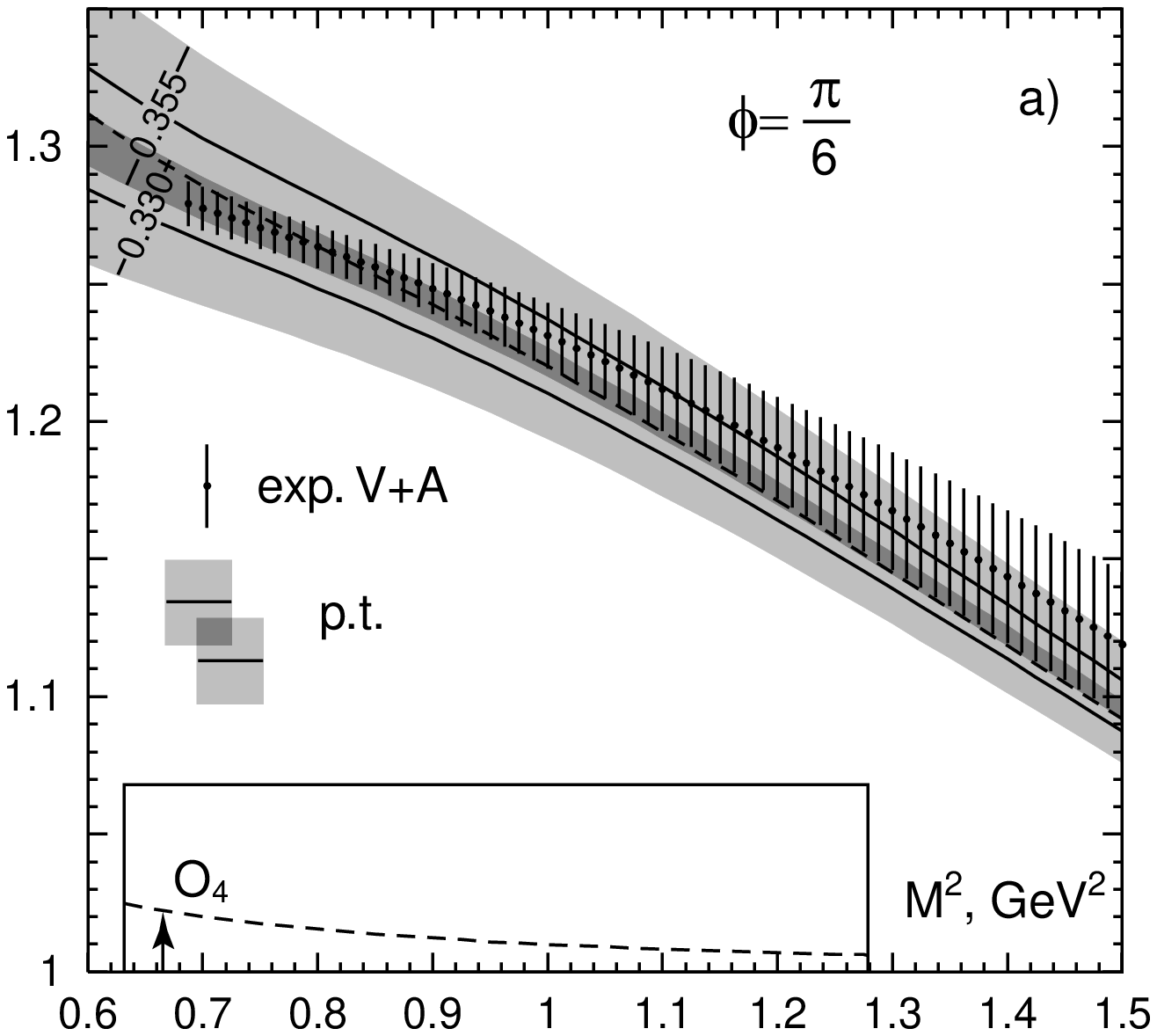} \hspace{5mm}
\includegraphics[width=80mm]{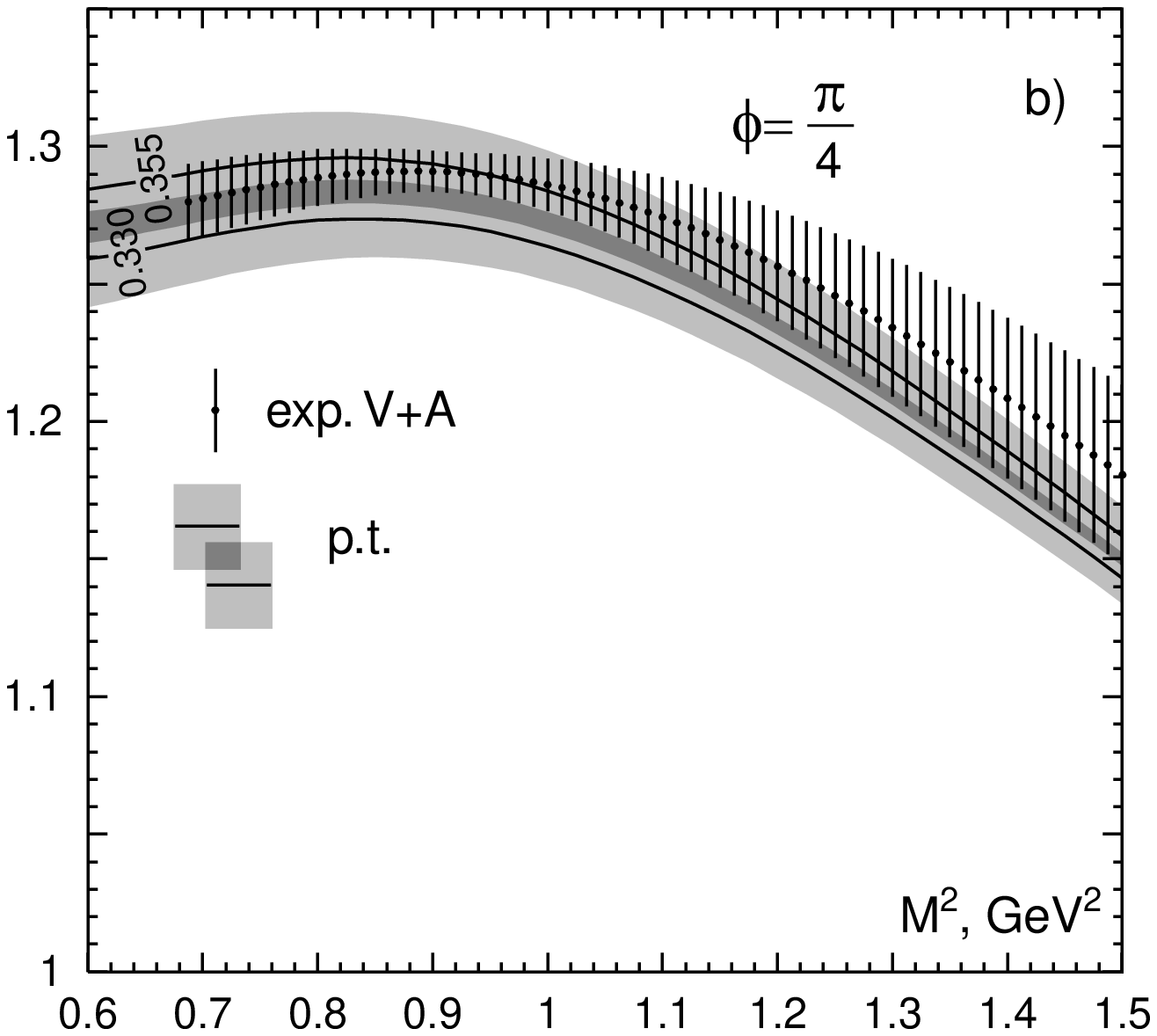}
\caption{Sum rule (\ref{rebor}) with $\phi=\pi/6$ (a) and $\phi=\pi/4$ (b).
The dash line is the contribution of the gluonic condensate equal to 
the central value of (\ref{gc}) added to the 0.330-perturbative
curve.} 
\label{set1819}
\end{figure*}

Since $O_8^{V+A}$ is small,
we shall be concerned with the $D=4,6$ operators. They can be conveniently
separated by taking the real part of the Borel transform (\ref{borvpa})
with complex argument:
$$
{\rm Re}\,B_{\rm exp}(M^2e^{i\phi})={\rm Re}\,B_{\rm pt}(M^2e^{i\phi})
$$
\be
 +2\pi^2 \sum_{k\ge 2} {\cos{(k\phi)}\, O_{2k}^{V+A}\over (k-1)!\, M^{2k}}
\label{rebor}
\ee
At $\phi=\pi/6$ the operator $O_6$ disappears and at $\phi=\pi/4$
there is no $O_4$ in the r.h.s. Fig.~\ref{set1819}a,b displays both these
possibilities. Vertical bars correspond to ${\rm Re}\,B_{\rm exp}$, 
while the solid line with shaded area around it, labeled with ``0.355''
mark, shows purely perturbative contribution ${\rm Re}\, B_{\rm pt}$
computed with initial condition $\alpha_s(m_\tau)=0.355$. The same is done
for $\alpha_s(m_\tau^2)=0.330$, which is the lowest possible value within
the error range (\ref{alpha_f}) (for a reason which will become 
clear later).
 
Consider at first $\phi=\pi/4$. The gluonic condensate $\left<G^2\right>$,
and consequently $O_4^{V+A}$ must be positive. Therefore, the 
perturbative curve must go below experimental one, if the discrepancy
is explained by OPE. However Fig.~\ref{set1819}a shows, that 
the central value of 0.355-theoretical prediction goes above the experimental
band for $M^2<0.9 \, {\rm GeV}^2$. If we forget for a moment about 
theoretical uncertainty and assume that the theory should work 
for $M^2>0.6 \, {\rm GeV}^2$, as follows from our $V-A$ analysis,
this means that the condensate must be negative. 
It rather contradicts to our expectations.

But if we take slightly lower input $\alpha_s(m_\tau^2)$, the
perturbative curve will go down. The lowest possible value is
0.330, as follows from (\ref{alpha_f}). Indeed, in this case the central
theoretical value is below experimental bars. If one takes 
some point, say, $M^2=0.8\, {\rm GeV}^2$, then the following 
values of the gluonic condensate are acceptable:
$$
\left<{\alpha_s\over \pi}G_{\mu\nu}^aG_{\mu\nu}^a\right>=
0.006\pm 0.012 \, {\rm GeV}^4 \hspace{15mm}
$$
\be
\label{gc}
\mbox{for} \quad \alpha_s(m_\tau^2)=0.330 \quad
\mbox{and} \quad M^2>0.8\,{\rm GeV}^2
\ee   
Theoretical and experimental errors are added together. 
In principle our result (\ref{gc}) does not contradict SVZ
value. However, in order to achieve it, we must sit at ``the very edge
of errors'', which seems unlikely. 

Now we turn to $\phi=\pi/4$, Fig.~\ref{set1819}b. If $O_6^{V+A}$ is
positive, as OPE+factorization predict (\ref{o6vpa}), its contribution
to the r.h.s. of (\ref{rebor}) must be negative.  
This however strongly disfavors our previous choice
$\alpha_s(m_\tau^2)=0.330$, motivated by the sign of the gluonic condensate.
So it seems rather difficult to make 
both $\left<G^2\right>$ and $D=6$ operator positive simultaneously. 
Here the OPE predictions are in certain disbalance with the data.

This is not however a serious disagreement. First, both $O_4^{V+A}$
and $O_6^{V+A}$ are small enough. Available theoretical and experimental
accuracy is about $2-3\%$, which is not sufficient to specify the
values of both operators or to say something definite about their signs.

Second, these operators are not rigorously defined objects in 
perturbation theory. We do not have an algorithm to find their values
from the first principles. So any statement about their properties
should be considered with care. Moreover, it is not clear whether
one could define them independently of the perturbative series,
to which they are added. In fact, different prescriptions
to separate the so-called perturbative and nonperturbative terms may lead
to different results, at least at the level of 1\%.

\section{summary}

\begin{enumerate}
\item
The $V-A$ polarization operator $\Pi^{(1)}_{V-A}(Q^2)$ is well described 
by OPE series for $Q^2\ge 1\,{\rm GeV}^2$.
\item
The operator $O_6^{V-A}$ is approximately 2 times larger than what
expected from QCD and low energy theorems. 
\item
The $V+A$ polarization operator $\Pi^{(1+0)}_{V+A}(Q^2)$ is well described
by purely perturbative terms for $Q^2>1\,{\rm GeV}^2$.
\item
Current theoretical and experimental accuracy is not sufficient to 
determine the value of the gluonic condensate $\left<G^2\right>$. However
it is likely to be much lower than commonly accepted SVZ value.  
\end{enumerate}

\section*{Acknowledgment}
Author thanks SLAC group for kind hospitality.
Work supported in part by Award No RP2-2247 of US Civilian Research and
Development Foundation for Independent States of Former Soviet Union
(CRDF), by Russian Found
 of Basic Research grant 00-02-17808 and INTAS Call 2000, project 587.


\begin{thebibliography}{99}
\bibitem{ALEPH2}
ALEPH collaboration: R. Barate et al, Eur. J. Phys. {\bf C4} (1998) 409
\bibitem{OPAL}
OPAL collaboration: K. Ackerstaff et al, Eur. J. Phys. {\bf C7} (1999) 571
\bibitem{IZ}
B.L. Ioffe and K.N. Zyablyuk, Nucl. Phys. {\bf A687} (2000) 437
\bibitem{GIZ}
B.V. Geshkenbein, B.L. Ioffe and K.N. Zyablyuk, hep-ph/0104048
\bibitem{SVZ}
M.A. Shifman, A.I. Vainstein and V.I. Zakharov, 
Nucl. Phys. {\bf B147} (1979) 385
\bibitem{BI}
V.M. Belyaev and B.L. Ioffe, Sov.Phys. JETP {\bf 56} (1982) 493
\bibitem{AC}
L.-E. Adam and K.G. Chetyrkin, Phys. Lett {\bf B329} (1994) 129
\bibitem{RVL}
T. van Ritbergen, J.A.M. Vermaseren and S.A. Larin, 
Phys. Lett. {\bf B400} (1997) 379
\bibitem{SS}
L.R. Surgaladze, M.A. Samuel, 
Phys. Rev. Lett. {\bf 66} (1990) 560, e. ibid. 2416;
S.G. Gorishny, A.L. Kataev and S.A. Larin, Phys. Lett. {\rm B259} (1991) 144
\bibitem{ShSol1}
D.V. Shirkov and I.L. Solovtsov, Phys. Rev. Lett. {\bf 79} (1997) 1204
\bibitem{BNP}
E. Braaten, S. Narison and A. Pich, Nucl. Phys. {\bf B373} (1992) 581
\end{thebibliography}
\end{document}